\documentclass[aps,twocolumn,prl]{revtex4-1} 
\usepackage[english]{babel}
\usepackage{amsmath}

\usepackage{color}

\usepackage{amsthm}
 
\usepackage{url}
\usepackage{latexsym}
\usepackage{amsfonts,amssymb,amstext}

\usepackage{graphicx}
\usepackage{enumerate}

\theoremstyle{plain}

\theoremstyle{definition}

\theoremstyle{remark}

\newcommand{\cS}{{\mathcal S}}
\newcommand{\cX}{{\mathcal X}}
\newcommand{\cM}{{\mathcal M}}
\newcommand{\cP}{{\mathcal P}}

\newcommand{\RR}{{\mathbb R}}

\newcommand{\ra}{{\rightarrow}}
\newcommand{\ve}{\varepsilon}
\newcommand{\De}{\Delta}
\newcommand{\ignore}[1]{{}}

\newcommand{\MarkovKernel}{\mathcal{P}}

\newcommand{\Q}{\mathbb{Q}}

\newcommand{\SP}{\mathbf {SPACE}}
\newcommand{\TP}{\mathbf {P}}
\newcommand{\NP}{\mathbf {NP}}
\newcommand{\BQP}{\mathbf {BQP}}
\newcommand{\PSP}{\mathbf {PSPACE}}

\newcommand{\BQPSP}{\mathbf {BQPSPACE}}
\newcommand{\NPSP}{\mathbf {NPSPACE}}
\newcommand{\LSP}{\mathbf {LOGSPACE}}

\newcommand{\randomap}{\mathcal{S}}

\newcommand{\rev}[1]{\textcolor{black}{#1}}
\newcommand{\mrev}[1]{\textcolor{black}{#1}}
\newcommand{\crev}[1]{\textcolor{black}{#1}}
\newcommand{\mrevtwo}[1]{\textcolor{black}{#1}}

\newcommand{\SBCT}{{\bf SBCT }}
\newcommand{\SA}{{\bf SA }}
\newcommand{\SBCTnospace}{{\bf SBCT}}
\newcommand{\SAnospace}{{\bf SA}}

\begin{document}

\title{Space-bounded Church-Turing thesis and computational tractability of closed systems}
\author{Mark Braverman}
\affiliation{Computer Science Department, Princeton University}
\author{Crist\'obal Rojas}
\affiliation{Departamento de Matem\'aticas, Universidad Andres Bello}
\author{Jonathan Schneider}
\affiliation{Computer Science Department, Princeton University}

\date{\today}


\begin{abstract}We report a new limitation on the ability of physical systems to perform computation -- one that is based on generalizing 
the notion of memory, or storage space, available to the system to perform the computation. Roughly, we define memory as the maximal amount of information that the evolving system can carry from one instant to the next. We show that memory is a limiting factor 
in computation even in lieu of {\em any} time limitations on the evolving system --- such as when considering its equilibrium 
regime.  We call this limitation the Space-Bounded Church Turing Thesis (\SBCTnospace). \mrevtwo{ The \SBCT is supported by a Simulation Assertion (\SAnospace), which states that predicting the long-term behavior of bounded-memory systems is computationally tractable. In particular, one corollary of \SA is an explicit bound on the computational hardness of the long-term behavior of a discrete-time finite-dimensional dynamical system that is affected by noise. We prove such a bound explicitly. }
\end{abstract}

\maketitle



Can we use computers to predict the future of evolving physical systems?  What are the computational capabilities of physical systems?   The fundamental Church-Turing thesis (CTT) \cite{turing1939systems}, and its physical counterparts \cite{Gandy80, Copeland07}  assert that any computation that can be carried out in finite time by a physical device can be carried out by a Turing Machine. 
 The thesis \rev{is sometimes} paraphrased in the following way: provided all the initial conditions with arbitrarily good precision, and
random bits when necessary, the Turing Machine can \emph{simulate} the physical
system $S$ over any fixed period of time $[0,T]$ for $T<\infty$. 

\rev{However, there exists conceivable situations which, while respecting all physical principles,  would  allow for nature  to exhibit behavior that cannot be simulated by computers \cite{nielsen1997computable, kieu2003quantum}. Note that the power of a physical process which is being used as a computer,  critically depends on our ability  to prepare the system and take measurements of it. Therefore, the impossibility to simulate some natural processes  does not immediately  contradict CTT.  In particular, it is not clear that the finite state spaces accessed by (quantum or classical) computers are sufficient to simulate, with arbitrary accuracy, all the processes one finds in nature, which may take place in infinite-dimensional  spaces \cite{nielsen1997computable}}.

Moreover, even if we can simulate a system for any fixed period of time $T$, in many situations one would like to know more and predict the  asymptotic properties of the system as $T\to \infty$, i.e. as it reaches its equilibrium regime.   \crev{In this case}, the computational unsolvability of problems like the Halting Problem --- itself a long-term property of Turing Machines --- implies that rich enough physical systems may exhibit non-computable asymptotic behavior \cite{Neumann66,Wolfram:84,Minsky:67,Sontag:91,Fredkin:2002,Mo91,Koiran:94,Reif:1994,AsMalAm95,BY06,BraYam07,GalHoyRoj07c}. 

\rev{As Feynman describes it \cite{feynman1982simulating}, to simulate the statistical asymptotic behavior of a physical system (say its equilibrium regime) means to have a  machine which,  when provided with a sequence of uniform random bits as input, outputs a sequence of states of the system \emph{with exactly the same} probability as nature does.  Note that even if there is a finite number of states which are \emph{distinguishable} for the physical measurement, the associated probability distribution may well be \emph{continuous}. The  non-computable examples mean that this infinite time horizon simulation is sometimes just not possible. For instance, \crev{there exist computable dynamical systems (e.g. maps on the unit interval \cite{GalHoyRoj07c} or cellular automata \cite{HellouinSablik}) for which there is a positive measure set of initial conditions leading to the same equilibrium regime --- so it is a ``physical state'' --- that yet no Turing machine can simulate in this way.}} 

On the other hand,  it has also been observed that this analysis may be affected by restricting some of the features related to the physical plausibility of the systems considered, such as  dimensionality, compactness, smoothness or robustness to noise --- the long term behavior of such restricted systems  may be easier to predict \cite{Mo91, nielsen1997computable, Navot04, BY06,BraYam07}.  

In this Letter, we report a new \crev{bound on the ability of physical systems to perform computation} --- one that is based on generalizing 
the notion of storage space from computational complexity theory to continuous physical systems. More precisely, \crev{we provide a formal definition of memory for physical systems and postulate an explicit quantitative bound on the computational complexity of their simulations. According to our postulate, bounded memory physical systems should not exhibit non-computable phenomena even in the infinite-time horizon. As evidence for our postulate, we rigorously prove that for compact noisy systems,  the non-computable phenomenon is broken by the noise even in the infinite-dimensional case. } \mrev{Moreover, to substantiate the quantitative part of the thesis, we show that if the noise is not a source of additional complexity, then the additional space requirements for simulating the system below the noise threshold are minimal.} 


Consider a   closed, stochastic system $\cS=X_t$ over a state space $\cX$. 
If the time $t$ is discrete,   define the {\em memory available to $\cS$ } as
\begin{equation}
\label{eq:mem}
\cM(\cS):= \sup_t \sup_{\text{$\mu$ distribution on $\cX$}} I_{X_t\sim \mu}(X_t;X_{t+1}). 
\end{equation} 
Here $I(X_t;X_{t+1})$ is Shannon's mutual information \cite{CoverT91}. If $f(x,y)$ is the PDF of the distribution of $(X_t,X_{t+1})$
where $X_t\sim \mu$ and $X_{t+1}\sim X_{t+1}|_{X_t}$, then 
\begin{equation}
\label{eq:info}
 I_{X_t\sim \mu}(X_t;X_{t+1}) := \iint f(x,y) \log \frac{f(x,y)}{f(x)f(y)}\, dx\, dy.
\end{equation} 
We take the supremum over all possible distributions $\mu$. Therefore $\cM(\cS)$ measures the maximum amount of information 
the system can carry from one time step to the next. Note that if the space $\cX$ is finite of size $N$ then $\cM(\cS)$ is bounded 
by the entropy $H(X_t)\le \log_2 |\cX|=\log_2 N$. As discussed below, in the presence of noise, all bounded finite-dimensional 
systems have finite memory available.

For continuous-time systems we define memory available at time lapse $\De t$ as the amount of information that may be preserved 
for $\De t$ time units:
\begin{equation}
\label{eq:memDt}
\cM_{\De t}(\cS):= \sup_t \sup_{\text{$\mu$ distribution on $\cX$}} I_{X_t\sim \mu}(X_t;X_{t+\De t}). 
\end{equation} 
Information theoretic considerations imply that $\cM_{\De t}(\cS)$ is a non-increasing function of $\De t$. 
The time-lapse $\De t$ is chosen to be the highest among the values of $\De t$ for which the behavior of the system at time 
scales below $\De t$ is dynamically and computationally simple. 
It is possible to artificially construct an example where $\lim_{\De t\ra 0} \cM_{\De t}(\cS)=\infty$, and where by encoding computation on a shrinking set of time 
intervals the computational power of the system is unbounded \cite{copeland02}. However, it has been pointed out \cite{Lloyd01,Lloyd2002} that quantum mechanical considerations impose an ultimate lower bound $\De t\geq t^{*}$ \footnote{$t^{*} = h/4E$, where $h$ is Plank's constant and $E$ is the average energy of the system.} on the time  it takes for a physical device to perform one logical operation.


 We postulate that the memory $\cM(\cS)$ is an intrinsic limitation \rev{on the ability of physical systems to perform computation}.  We call the  limitation the Space-Bounded Church Turing thesis (\SBCT):

\smallskip
\noindent
{\bf[\SBCTnospace]:} {\em If a physical system $\cS$ has memory $s=\cM(\cS)$ available to it, \rev{then it is only capable of  performing computation  in the complexity class $\SP(s^{O(1)})$, even when provided with unlimited time.}}

\mrev{
\SBCT is supported by the following assertion:
}
\smallskip
\noindent
{\bf[Simulation Assertion, \SAnospace]:} \mrev{
{\em
The problem of simulating  the asymptotic behavior of a \crev{physical} system $\cS$ as in \SBCT with $n$ precision bits is in the complexity class $\SP\left((s +\log n)^{O(1)}\right)$.} }


\mrevtwo{
\SA implies, in particular,  that the long-term 
behavior of bounded-memory systems is computable.
This covers a broad class of noisy  systems. Interestingly, a number of low-dimensional systems with non-computable  long-term behavior is known \cite{Mo91,Koiran:94,Reif:1994,AsMalAm95,BY06,BraYam07,GalHoyRoj07c}. These examples require considerable care in their construction. As explained below, assuming the \SBCT one should expect these constructions to be delicate, to the point of making them physically implausible.} 

\mrevtwo{
It is clear that \SA implies \SBCTnospace. While, logically speaking, the converse also (almost) holds, it is still useful to make a distinction between the two statements. A low-memory system $\cS$ may be hard to simulate, for example, because of the hardness of   the noise operator. Such a system would violate \SAnospace. However, it might still essentially satisfy \SBCT --- being incapable of performing computation outside the
class  $\SP(s^{O(1)})$ --- save for the problem of simulating $\cS$ itself.
}


\SBCT can be considered in the context of other \mrev{quantitative} variants of the Church-Turing Thesis, \mrev{notably} the Extended Church-Turing Thesis (ECT) which asserts that 
physically-feasible computations are not only computable, but are {\em efficiently computable} in the sense of computational 
complexity theory \cite{Parberry86}.  Whereas previous discussions of efficiency focused on {\em time complexity} \cite{vergis86, siegel98, siegel99, abrams98, aarons05}, we shift the discussion 
to {\em storage space complexity} (known as space complexity in the Computer Science literature). This shift has the benefit of 
allowing one to make assertions bounding the computational power of systems even when provided with unlimited time ---  we e.g. can allow the system to reach equilibrium at $t\ra\infty$, and consider the outcome to be the output of the computation. We assert that this outcome will still not enhance the computational power of the system beyond its memory constraints.


 In the theory of computational complexity, $\SP(S(n))$ is the complexity class of problems that can be solved
by a Turing Machine which uses at most $S(n)$ bits of memory to solve instances of size $n$ \cite{sipser2006introduction,arora2009computational}. Of particular interest
are the classes of problems $\PSP$ and $\LSP$ where $S(n)=n^{O(1)}$ and $S(n)=O(\log n)$, respectively \footnote{We recall that the notations $f=O(g)$ and $f=\Theta(g)$  mean that, up to a multiplicative constant,  $f$ is bounded by $g$ and  $f$ is the same as $g$, respectively.}. Putting these classes in the context
of $\TP$ and $\NP$, the following chain of inclusions is known:
$$
\LSP\subset\TP\subset\NP\subset\PSP. 
$$
All of these inclusions are believed to be strict, although only the fact that $\LSP\subsetneq\PSP$ is known. 

Space-bounded complexity classes exhibit several important robustness properties that do not have 
a parallel when considering time-bounded computation. For example, the space-bounded analogue 
of $\TP\stackrel{?}{=}\NP$ has been resolved in the affirmative: $\PSP=\NPSP$ \cite{savitch1970relationships} --- thus 
$\PSP$ is closed under the use of non-determinism. The question of whether quantum computation 
speeds up computation time in some cases, i.e. whether $\TP\subsetneq\BQP$, remains open, but 
likely the answer is that it does \cite{shor1997polynomial,nielsen2000quantum}. In the case of {\em space} limitations, it is known that 
$\BQPSP=\PSP$, and thus quantum computing is not particularly useful \cite{watrous1999space} (suggesting that, unlike the ECT, the \SBCT has a good 
chance of holding in a quantum world). 

A bound of $S(n)$ on the amount of memory used by a computation means that the machine may be in 
at most $2^{S(n)}$ distinct states. If the computation is deterministic, this imposes a natural hard limit of $2^{S(n)}$ on 
its computation time: the computation either terminates in $2^{S(n)}$ steps, or ends up in an infinite loop. 
If the computation is randomized, then it naturally translates into a Markov chain on its $2^{S(n)}$ states. 
The stationary distribution(s) of the chain, which can be computed \mrev{in $\operatorname*{poly}({S(n)})$ space}, characterize the infinite-time horizon behavior 
of the machine. We assert that more generally, the ability of physical systems to remember information is the limiting 
factor for their computational power.  


While in many  cases the complexity of the system falls below 
the bound provided by \SBCTnospace, the power of \SBCT partially arises from the fact that it is generally much easier to estimate the memory available to a system than its computational power/hardness.

The non-computability constructions \mrev{mentioned earlier} mean that while analytic methods can prove {\em some} long-term properties of {\em some} dynamical systems, for ``rich enough" systems, one cannot hope to have a general closed-form {\em analytic} algorithm, i.e. one that is not based on simulations,  that computes the properties of its long-term behavior. This  fundamental phenomenon  is qualitatively different from chaotic behavior, and has even led some researchers to claim \cite{Wol02} that the enterprise of theoretical physics itself is doomed from the outset; rather than attempting to construct solvable mathematical models of physical processes,
computational models should be built, explored, and empirically analyzed.

However,  it is a notable fact that in all the specific low-dimensional examples the non-computability phenomenon is {\em not robust to noise}: all these constructions are based on a fine structure responsible for Turing simulation which is destroyed once one introduces even a small amount of noise into the system. This has been explicitly observed e.g. for neural networks \cite{Maass97} and reachability problems  \cite{Asa01}. This is consistent with the \SBCTnospace: a low-dimensional compact system affected by noise 
becomes a bounded-memory system, and is therefore explicitly limited in its computational power, and cannot serve as a universal computer. 

In \cite{Graca11}, an interesting example of a constant-dimensional analytic system capable of robustly performing universal computation is constructed.  However, this system acts on an unbounded domain, and has therefore infinitely many robustly distinguishable states; i.e. infinite memory. This is again consistent with the \SBCTnospace.


\crev{We now turn to the rigorous analysis of } discrete-time dynamical systems over continuous spaces, affected by random noise. In such models, the evolution is governed by a deterministic map $T$ acting on phase space $\cX$,  together with a small random noise $p^{\ve}$. The noisy system $\cS_{\ve}$  jumps, in one unit of time,  from state $x$ to $T(x)$ and then disperses randomly  around $T(x)$ with  distribution $p^{\ve}_{T(x)}$. The 
parameter $\ve$ controls the ``magnitude" of the noise, so that  $p^{\ve}_{T(x)}(\cdot)\to T(x)$ as $\ve\to 0$ \cite{Kif88}. For example, $p^{\ve}_{T(x)}(\cdot)$ could be taken to be uniform on an $\ve$-ball around $T(x)$ or a Gaussian with mean $T(x)$ and variance $\ve$.  In all what follows we will assume, for the sake of simplicity, that the underlying system is one-dimensional and size($\cX$)=1. That is, $\cX$ can be thought of 
as the interval $[0,1]$.

By expressing mutual information in terms of entropy and conditional entropy, it is not hard to estimate the memory of the system $\cS_{\ve}$ for each of these types of noise (uniform on an $\ve$-ball  or Gaussian). Indeed, if $f_{X}$ stands for the PDF of a random variable $X$, then the entropy of $X$ is defined by
$$
H(X)=- \int f_{X}(x)\log(f_{X}(x))\,dx,
$$
and mutual information can be expressed as
$$
I(X_t;X_{t+1}) = H(X_{t+1})-H(X_{t+1}|X_{t}).
$$

On the one hand, since $H(p^{\ve})= \Theta(\log(\ve))$ for both uniform on an $\ve$-ball and Gaussian distributions and since $H(X_{t+1})\le 0$ and $X_{t+1}|X_{t}\sim  p^{\ve}$, 
we obtain that $I(X_{t+1};X_t)\le O(\log 1/\ve)$. On the other hand, $H(X_{t+1})$ is maximized by the uniform distribution on $\cX$, having a value of $\log(\text{size}({\cX}))=0$.  It follows that $I(X_t;X_{t+1})$ is maximized by this distribution as well, and therefore $\cM(\cS_{\ve})=
\Theta(\log(1/\ve))$. 
The \SBCT  then predicts that the computational power of the system $\cS_{\ve}$ is in the complexity class $\SP\left(\log^{O(1)}(1/\ve)\right)$. 

How can the actual computational power of  these systems be estimated?  In order to give an upper bound one would have to give a generic algorithm for the noisy system  that computes  its long-term features. This would establish the \SA for the system, and thus imply the \SBCTnospace. In order to give a lower bound one would have to show that even in the presence of noise the system is capable of simulating a Turing Machine subject to memory restrictions.  We now explain how to prove such bounds. 

 Since the evolution of these systems is stochastic, only the statistical properties can be studied ---  instead of asking whether the system will ever fall in a given region $B$, we shall ask what is the probability of the system being in such a region, as $t\ra\infty$. 

These properties  are mathematically described by the {\em invariant measures} of the system --- the possible statistical 
 behaviors  once the system has converged to a ``steady state" distribution. Quantities such as Lyapunov exponents or escape rates can be computed from the relevant invariant measure. Standard references on this material are \cite{Wal82, Pet83,Man87}.
 
 Here,  by computing a probability distribution $\mu$ over $[0,1]$ we mean to have a finite algorithm $A$ that can produce arbitrarily good rational approximations to the probability of any interval with rational endpoints. That is, the algorithm $A$, upon input $(a,b,\delta)\in \Q^{3}$, must output a rational number $A(a,b,\delta)$ satisfying $|A(a,b,\delta) -  \mu[a,b]|\leq \delta$. See for instance \cite{HoyRoj07}. This definition is equivalent to the existence of a probabilistic machine producing a sequence of states distributed exactly according to $\mu$ \cite{knuth1976complexity}. 
 
\crev{Our first result, which can be seen as supporting the qualitative part of \SBCTnospace,} shows that the addition of any amount of noise to a system is sufficient to destroy any non-computable behavior, even in the infinite-dimensional case. 



\medskip
\noindent {\bf Statement~A:} {\em  If a compact system is affected by small random $\ve$-noise as described above, then all its ergodic invariant measures are computable.}
\medskip

 Intuitively,  this theorem  says noise turns asymptotic statistical properties from non-computable to computable. 
Its proof essentially follows from the fact that the presence of noise forces the system to have only ``well separated'' ergodic measures. An exhaustive search can then be performed, and compactness guarantees that all such measures will be eventually found (there can only  be finitely many of them). We note that the result holds even if the state space is infinite dimensional. We refer to \cite{Braver12} for a complete proof. 



Thus, we know that in presence of $\ve$-noise, ergodic measures are all computable. In addition, according to the \SBCTnospace, their computational power should be bounded in terms of their dimension and size.
In order to give an upper bound, we prove a version of the \SA by exhibiting an algorithm that computes the invariant measure to arbitrary accuracy
using very little space. Specifically, we show:

\smallskip
\noindent {\bf Statement~B:} {\em Let $\cS$ be a compact, constant-dimensional system affected by $\ve$-Gaussian noise. Suppose that the transition function $f$ is 
uniformly analytic and can be computed to within precision $2^{-m}$ using $O(\log m)$ space. Then the invariant measure of the noisy system $S_\ve$
can be computed with a given precision $2^{-n}$ in $\mathbf{SPACE}(\operatorname*{poly}(\log 1/\ve)+\operatorname*{poly}(\log n))$. 
}
\medskip

This statement implies, in particular, that the long-term behavior of noisy systems at scales below the noise level  is computable in time quasipolynomial in $n$. \mrevtwo{Intuitively, this means that, at the right scale,  the behavior of the system is governed by the efficiently predictable micro-analytic structure of the
noise, rather than by the macro-dynamic structure of the system that can be computationally difficult to predict. 
}


The formal proof of the above statement can be found in the accompanying paper \cite{SharperBounds}. Moreover, up to the polynomial factors, the statement can be shown to be tight: we can robustly separate $1/\ve$ states of $\cS_\ve$, and thus simulate a computation that uses $\sim\log 1/\ve$ bits of memory. Therefore, simulation using less than $\log 1/\ve$ bits of memory is impossible due to the Space Hierarchy Theorems \cite{arora2009computational}. Note that the output of a precision-$2^{-n}$ calculation requires $\ge n$ bits to write down. In the context of space-bounded computation, the output is stored in a write-only memory that is not part of the computation space. Still, in order to be able to write to a size-$n$ outside memory, one needs to at least store indexes using $\log n$ bits, and thus the dependence on $n$ is also optimal up to polynomial factors. 

The algorithm establishing Statement~B and its analysis consists of two main parts. The first idea is to exploit the mixing properties of the transition operator $\MarkovKernel$ of the perturbed system $\randomap_{\ve}$. The transition operator contains Gaussian noise, and it thus has a spectral gap of at least $\exp(-1/\ve^2)$, and will mix in time on the order of $T\approx \exp(-1/\ve^2)$. We represent density functions of measures using piece-wise analytic functions with each piece of size $\approx \ve$. On each piece we approximate the corresponding analytic function using $\approx n$ terms of its Taylor expansion, so that the density function is represented by a point in $\RR^D$ where $D\approx n/\ve$.  When we consider the action of the transition operator $\MarkovKernel$ on these coefficients, we obtain a linear map $\cM_\cP$ whose coefficients can be computed in space $\operatorname*{poly}(\log 1/\ve)+\operatorname*{poly}(\log n)$. By the mixing property, to approximate the invariant measure of $\cP$ it suffices to raise 
$\cM_\cP$ to the $T$-th power. 

The second part of the argument deals with raising a $D\times D$ matrix $\cM_\cP$ to power $T\approx 2^D$ using only $\operatorname*{poly}(\log D)$ space. To the best of our knowledge, this problem has been previously addressed when $T$ is polynomial but not exponential in $D$. The proof in \cite{SharperBounds} uses a number of  techniques in space-efficient computation to obtain a degree-$O(D)$ polynomial $p(\cdot)$ such that the entries of $p(\cM_\cP) -  \cM_\cP^T$ have magnitude $\leq 2^{-n}$.

In conclusion, we postulated a principle that allows us to quantitatively bound the  computational power of any device built out of a closed physical system --- even when the device is allowed to run for an unlimited amount of time --- in terms of the memory of the system. We have shown that this bound is tight for systems modeled by randomly perturbed dynamical processes, which account for a large part of physics. Additionally, we have shown that the asymptotic behavior of these systems can be computed at arbitrary precision, and that when computing below the noise level, the simulation can be achieved using an extremely limited amount of memory. Concerning quantum systems, the fact that general models like Topological Field Theories can be efficiently simulated by quantum computers \cite{freedman2002} which, in turn, can be simulated by classical ones with only a quadratic increase in memory \cite{watrous1999space}, suggests that our results apply in the quantum world as well.

\subsection*{Acknowledgments}
MB was partially supported by an Alfred P. Sloan Fellowship, an NSF CAREER award (CCF-1149888), NSF CCF-0832797,  a Turing Centenary Fellowship, and a Packard Fellowship in Science and Engineering. CR was partially supported by projects Fondecyt 1150222,  DI-782-15/R Universidad Andr\'es Bello and Basal PFB-03 CMM-Universidad de Chile. 

We would like to thank  Jos\'{e} Aliste, Eric Allender, Cameron Freer, Andy Gomberoff, Aram Harrow, Avinatan Hassidim, Giorgio Krstulovic, and the anonymous referees for their advice and comments on earlier versions of the manuscript.

\bibliography{biblioPRL}

\end{document}